\documentclass[manuscript]{acmart}
\usepackage{multicol}
\usepackage{tikz}

\AtBeginDocument{%
  }

\setcopyright{acmlicensed}
\copyrightyear{2025}
\acmYear{2025}
\acmDOI{XXXXXXX.XXXXXXX}
\acmConference[MobileHCI '25]{ACM MobileHCI International Conference on Mobile Human-Computer Interaction}
{September 22 - September 26 2025}{Sharm El-Sheikh, Egypt}

\acmISBN{978-1-4503-XXXX-X/2018/06}

\usepackage{fontawesome5}

\newcounter{implication}
\usepackage{xcolor} 

\newcommand{\summaryBox}[2]{%
  \refstepcounter{implication}%
  \par\medskip
  \noindent
  \begingroup
  \setlength{\fboxsep}{6pt}
  \setlength{\fboxrule}{0.25pt}
  \colorlet{framecolor}{black}
  \colorlet{titlecolor}{black}
  \colorlet{backcolor}{white}
  \fcolorbox{framecolor}{backcolor}{%
    \begin{minipage}{\dimexpr\linewidth-2\fboxsep-2\fboxrule\relax}
      \textbf{\textcolor{titlecolor}{Implication~\theimplication: #1}}
      #2
    \end{minipage}%
  }%
  \par\medskip
  \endgroup
}

\begin{document}

\setcopyright{acmlicensed}
\acmJournal{PACMHCI}
\acmYear{2025} \acmVolume{9} \acmNumber{5} \acmArticle{pn6865} \acmMonth{9}\acmDOI{10.1145/3743738}

\title[User Understanding of Privacy Permissions in Mobile Augmented Reality]{User Understanding of Privacy Permissions in Mobile Augmented Reality: Perceptions and Misconceptions}

\author{Viktorija Paneva}
\email{viktorija.paneva@ifi.lmu.de}
\orcid{0000-0002-5152-3077}
\affiliation{%
  \institution{LMU Munich}
  \city{Munich}
  \country{Germany}}

\author{Verena Winterhalter}
\email{verena.winterhalter@ifi.lmu.de}
\orcid{0000-0003-0752-3480}
\affiliation{%
  \institution{LMU Munich}
  \city{Munich}
  \country{Germany}}

\author{Franziska Augustinowski}
\email{f.augustinowski@campus.lmu.de}
\orcid{0009-0004-7875-722X}
\affiliation{%
  \institution{LMU Munich}
  \city{Munich}
  \country{Germany}}

\author{Florian Alt}
\orcid{0000-0001-8354-2195}
\affiliation{%
  \institution{LMU Munich}
  \city{Munich}
  \country{Germany}}
\affiliation{%
  \institution{University of the Bundeswehr}
  \city{Munich}
  \country{Germany}}
\email{florian.alt@ifi.lmu.de}

\renewcommand{\shortauthors}{Paneva et al.}

\begin{abstract}
Mobile Augmented Reality (AR) applications leverage various sensors to provide immersive user experiences. However, their reliance on diverse data sources introduces significant privacy challenges. 
This paper investigates user perceptions and understanding of privacy permissions in mobile AR apps through an analysis of existing applications and an online survey of {120} participants. 
Findings reveal common misconceptions, including confusion about how permissions relate to specific AR functionalities (e.g., location and measurement of physical distances), and misinterpretations of permission labels (e.g., conflating camera and gallery access).
We identify a set of actionable implications for designing more usable and transparent privacy mechanisms tailored to mobile AR technologies, including contextual explanations, modular permission requests, and clearer permission labels. These findings offer actionable guidance for developers, researchers, and policymakers working to enhance privacy frameworks in mobile AR.
\end{abstract}


\begin{teaserfigure}
  \includegraphics[width=\textwidth]{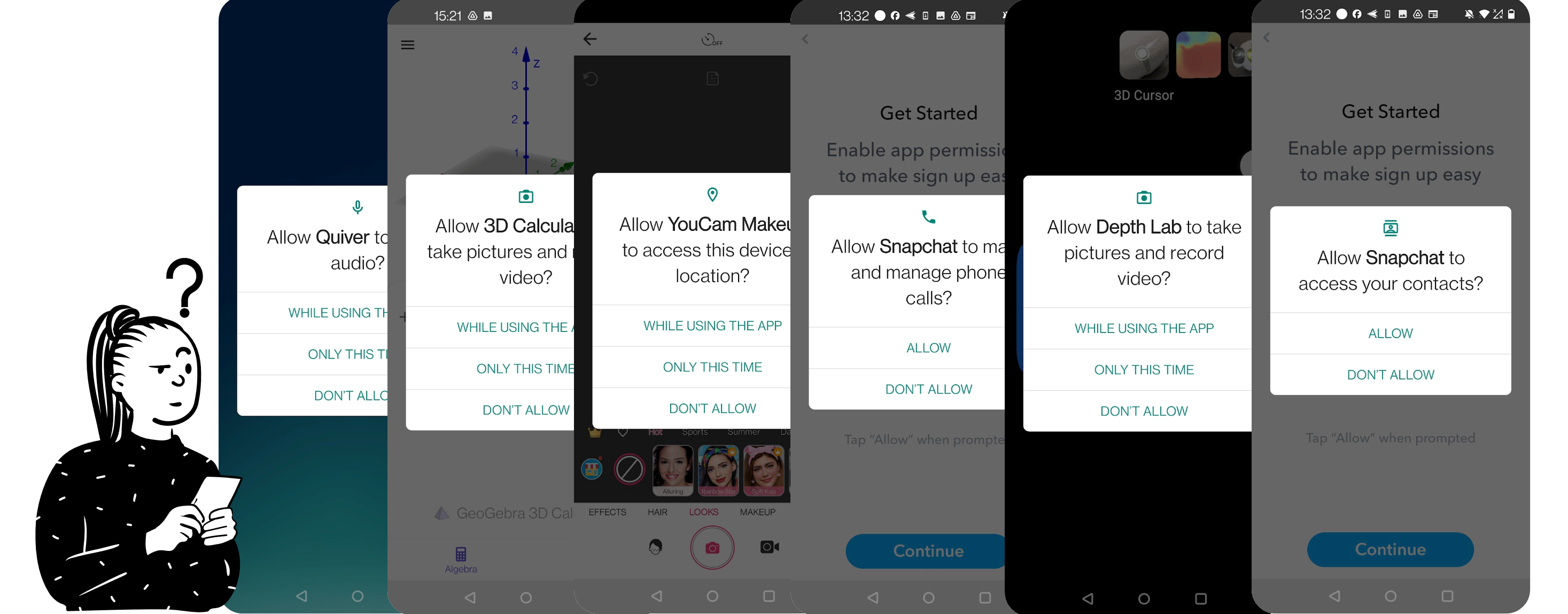}
  \caption{In this paper, we examine permission practices across popular mobile AR applications and explore user perceptions and misconceptions, identifying potential areas for improved transparency and understanding. (Apps accessed: September 2024)}
  \Description{}
  \label{fig:teaser}
\end{teaserfigure}


\maketitle

\section{Introduction}

Augmented Reality (AR) has emerged as a prominent tool for gaming, navigation, education, and other interactive domains, largely due to its ability to project digital elements onto real-world settings in real time \cite{harborth2021investigating}. Unlike typical smartphone applications, AR relies on continuous sensor fusion—combining camera streams, GPS data, motion tracking, and sometimes audio input—to generate immersive experiences. This sensor-rich environment creates a more nuanced privacy landscape: data are not only captured at discrete moments but can be collected persistently, potentially exposing bystanders, personal surroundings, and fine-grained behavioral patterns. These risks pose challenges for how consent is obtained and understood, particularly since current “notice and choice” frameworks often present simple yes/no prompts, neither conveying the intricacy of AR data gathering nor clarifying how this data is processed or shared \cite{sloan2014beyond}.

Recent privacy research in HCI has highlighted that people frequently accept permissions without fully understanding the implications or decline them outright out of distrust, both of which can undermine a meaningful user experience \cite{rothchild2017against}. In AR, these tensions are amplified because permissions commonly labeled as “camera” or “location” may allow complex tasks like real-time 3D mapping or continuous environment scanning. Equally concerning is that these apps may gather data about objects or individuals in a user’s vicinity, affecting privacy well beyond the primary user. Despite this complexity, developers continue to rely on generic system prompts with minimal tailoring for AR scenarios.

One avenue to address these gaps is through more explicit disclosures or just-in-time explanations that clarify AR’s sensor usage. For example, informing users that “camera access enables continuous environmental mapping, including background objects and people” may improve comprehension and reduce suspicion regarding how the application processes visual data. However, providing too many or overly detailed notices can result in fatigue, where users stop reading explanations entirely. Thus, striking a balance between clarity and brevity is essential. Moreover, future design approaches could include optional in-app tutorials or interactive demos that walk users through why the app needs certain permissions, making these rationales more transparent and actionable.

Our research is driven by the question: \emph{How are current mobile AR permissions perceived and understood by users?} 
We adopted a two-step methodology. First, we audited 23 widely used Android AR apps, documenting which permissions they request, how requests are timed (e.g., upon installation vs.\ just-in-time), and whether explanations were provided. 
Second, we conducted an online survey (N={120}) presenting participants with real-world permission prompts from our audit. By comparing actual permission requests with participant expectations and rationales, we identified persistent gaps—such as conflating camera and gallery permissions or misunderstanding how often location data might be accessed. Our findings highlight the need for more nuanced privacy mechanisms that reflect AR’s ongoing, sensor-intensive data capture and give users the means to make informed decisions about granting, limiting, or revoking permissions.

Looking ahead, as wearable AR devices become mainstream, privacy concerns may intensify. Head-mounted displays and real-time environment mapping can expose even more bystander data or sensitive spatial details. Traditional mobile consent models will likely prove insufficient for these scenarios, prompting further development of AR-specific permission policies and user education strategies. Ultimately, developing robust and user-centric privacy controls will be critical to sustaining public trust and maximizing the potential of AR technologies. 

\vspace{1mm}\noindent
\textbf{Contribution Statement.} We make three primary contributions. First, we conduct a systematic review of privacy permissions in a diverse set of mobile AR applications to show how sensor access is requested and (often insufficiently) justified. Second, we report findings from an online survey (N={120}) that sheds light on users’ perceptions, mental models, and misunderstandings when confronted with AR permission prompts. Third, we leverage these findings to identify design implications for more transparent, granular, and comprehensible AR permission mechanisms.


\section{Related Work}
This section reviews prior work on AR technologies, their adoption across industries, and emerging privacy considerations in these contexts.

\subsection{Augmented Reality and Mobile Adoption}
Augmented Reality (AR) is a ``technology that aims to digitally integrate and expand the physical environment or the user’s world, in real-time, by adding layers of digital information''~\cite{computers11020028}. 
Although the term ``Augmented Reality'' was first popularized by Caudell and Mizell~\cite{arth2015historymobileaugmentedreality,ARCaudellMizell}, early conceptual roots date back to the 1950s. 
Today, AR is applied in a wide array of domains, including gaming~\cite{FAQIH2022101958} (e.g., educational applications~\cite{EducationalARGames}), commercial settings~\cite{commercialIndustrialAR}, Industry~4.0~\cite{ARIndustry4.0}, tourism~\cite{SystemativReviewARTourism,ValueARTourism}, and medical procedures~\cite{ARMedicine}. 

Collectively, these studies underscore AR’s broad utility, but the underlying technology can vary in how digital content is anchored. On smartphones, AR implementations are typically classified as \emph{marker-based}, \emph{markerless}, or \emph{location-based}. Marker-based approaches detect predefined images or patterns (e.g., QR codes) to position virtual objects \cite{Boonbrahm2020,Paucher2010}, whereas markerless variants track environmental features in the camera feed \cite{Oufir2020}. Location-based AR uses GPS and other sensors to deliver context-aware overlays at specific coordinates \cite{Paavilainen2017}. Although dedicated headsets (e.g., Microsoft HoloLens) enable advanced AR functionality \cite{MicorsoftHololens}, cost and comfort constraints limit their broader adoption \cite{Harborth2021}. As a result, most consumer-oriented AR takes the form of mobile apps leveraging built-in cameras, motion sensors, and location data. Among high-profile examples, \emph{Pokémon Go} (launched in 2016) showcased mass-market AR potential by drawing large user numbers worldwide \cite{Harborth2021,Harborth2017,PokemonGoAR}.


\subsection{Privacy Concerns and Permission Practices {in Mobile and AR Contexts}}

{
Research on usable privacy in mobile applications highlights persistent usability challenges that limit user control, leading to privacy fatigue and privacy helplessness~\cite{Cho2021}.
Default privacy settings are often permissive and rarely adjusted, while options to modify them can be hard to locate or understand~\cite{LiteratureReviewUsbalePrivacy}. In response, various tools have been proposed to improve privacy usability. For example, SPARCLE aims to help users understand privacy policies through natural language input and visualizations~\cite{Brodie2005}, PriVs recommends personalized privacy permissions based on crowd-sourced behavior~\cite{Liu2018}, and SeePrivacy introduces contextual, in-app privacy policy prompts to improve readability and engagement~\cite{Pan2023}.}

While these tools suggest promising directions for improving privacy interactions in conventional mobile apps, they remain insufficient for mobile AR, which introduces a uniquely complex data environment.
On Android, permissions are categorized by risk level, with \emph{normal} permissions (e.g., setting time zones) granted automatically and \emph{dangerous} permissions (e.g., storage access) requiring explicit user consent~\cite{AnroidPermissionSystem,UserStudyAndroidPermissions}. Yet, no dedicated class of permissions covers AR-specific sensor data such as LiDAR or accelerometer outputs, even though these modalities can capture detailed spatial and environmental information~\cite{Harborth2021}. As a result, AR apps often bundle these functionalities under broader labels such as "camera" or "location", without conveying context about real-time object recognition or persistent environment mapping~\cite{harborth2021investigating}. 

This gap creates a mismatch between the advanced data capture typical of AR and the broad, non-AR-specific prompts presented to users~\cite{BystandersConsentAR}.
Users are thus left to interpret broad system prompts in highly technical contexts—often without enough information to make informed decisions. 


The literature on mobile AR is expanding but remains limited in its exploration of user-centric privacy challenges~\cite{Harborth2017}. Some studies point to unintentional surveillance and involuntary recording as pressing concerns~\cite{harborth2021investigating}, while others emphasize users’ desire for granular explanations of permissions and transparent opt-out mechanisms~\cite{BystandersConsentAR}. 
{However, what remains missing is a systematic understanding of how users actually perceive and interpret these permission requests in mobile AR. Without this baseline, it is difficult to design effective privacy solutions that match users’ mental models, address misconceptions, or support informed consent.}


\subsection{Summary}

Mobile AR applications depend on continuous, multimodal sensor input to deliver interactive experiences, yet current permission models do not distinguish AR-specific data from ordinary mobile app access. As a result, users encounter generic prompts that do not reflect the intensity, frequency, or implications of AR-specific data capture.

{Existing research has explored privacy usability in mobile contexts and raised concerns about AR’s potential to infringe on user and bystander privacy. However, little is known about how users actually interpret AR-related permission requests—what they expect, misunderstand, or mistrust. Without this understanding, privacy design for AR risks being misaligned with user needs and mental models.
}

{
Prior work has taken important steps toward characterizing user privacy concerns at scale. Nema et al.~\cite{Nema2022ICSE} , for example, analyzed user reviews to surface large-scale concerns about mobile app permissions. However, their method is limited to users willing to publicly post reviews, and it does not address the specific complexities of AR data practices.
Similarly, Gallardo et al.~\cite{Gallardo2023PETS} explored user concerns around AR glasses, offering valuable insights into perceptions of visible, wearable AR technology. Yet their participants were all confirmed AR users based in the United States. Our study instead targets a broader and more globally distributed population, including users who may not even be aware they are interacting with AR through their smartphones. This enables us to uncover a different class of misconceptions, especially among users encountering AR features embedded in apps like maps, games, or social media.}

{
Our work thus complements and extends these prior efforts by offering a focused, empirically grounded look at how users understand (or misunderstand) permission prompts in mobile AR contexts—where privacy risks are often high but poorly communicated. 
}

\section{Methodology}

This section details our multi-stage approach to studying privacy permissions in mobile AR. We first describe how we selected and audited 23 popular AR-enabled apps before outlining our design of an online survey. Finally, we discuss participant recruitment, data collection, and our analytical procedures for both quantitative and qualitative responses.

\begin{table*}[t]
    \caption{List of a selection of Android Smartphone Apps using Augmented Reality}
    \centering
    \footnotesize
    \begin{tabular}{l l l l l }    
 \toprule
  \textbf{Application} & \parbox{1.5cm}{\textbf{\# Downloads }\\\textbf{in Million}} &  \parbox{1.2cm}{\textbf{Type} \\ \textbf{of Choice}} &\parbox{1.2cm}{\textbf{ Timing of }\\ \textbf{Choice}}& \parbox{5cm}{\textbf{List of Permissions}} \\ 
 \midrule
 Google Maps & 10 000 & binary & at setup & camera, location \\ 
 \hline
 Instagram & 5 000 & multiple & just in time & \parbox{5cm}{location, contacts, camera, gallery, microphone}\\
 \hline
 Google Lens & 1 000 & multiple & just in time & camera, gallery\\
 \hline
 Google Translate & 1 000 & multiple & just in time & camera, gallery\\
 \hline
 Snapchat & 1 000 & multiple & at setup & \parbox{5cm}{contacts, microphone, camera, gallery, manage calls} \\ 
 \hline
 Pokémon Go & 500 & multiple & \parbox{2.5cm}{just in time/at setup} & \parbox{4cm}{contacts, camera, location}\\
 \hline
 YouCam Makeup & 100 & multiple & just in time & \parbox{5cm}{camera, microphone, location}\\
 \hline
 GIPHY & 50 & multiple & just in time & camera\\
 \hline
 AR Ruler App & 10 & binary & just in time & camera\\ 
 \hline
 Ikea Place & 10 & multiple & just in time & camera\\
 \hline
 Ingress Prime & 10 & multiple & at setup & location\\
 \hline
 Magicplan & 10 & multiple & just in time & camera\\
 \hline
 Stellarium Mobile & 10 & multiple & at setup & location\\ 
 \hline
 Foto Tattoo Simulator & 5 & multiple & \parbox{2.5cm}{just in time/at setup} & camera, gallery\\
 \hline
 SketchAR & 5 & multiple & just in time & camera\\
 \hline
 \parbox{3.2cm}{Assemblr Studio: Easy AR Maker} & 1 & multiple & just in time & camera\\
 \hline
 Augment - 3D AR & 1 & multiple & just in time & camera\\ 
 \hline
 GeoGebra 3D & 1 & multiple & just in time & camera\\
 \hline
 Meta Spark Player & 1 & multiple & at setup & \parbox{5cm}{camera, microphone}\\
 \hline
 \parbox{3cm}{Quiver - 3D Coloring App} & 1 & multiple & at setup & \parbox{5cm}{camera, microphone}\\
 \hline
 \parbox{3cm}{ROAR Augmented Reality App} & 0.1 & multiple & at setup & \parbox{5cm}{camera, microphone, location} \\ 
 \hline
 ARCore Depth Lab & 0.05 & multiple & at setup & camera\\
 \hline
 Vuforia Chalk & 0.05 & multiple & just in time & \parbox{2.5cm}{camera, microphone}\\
 \bottomrule
\end{tabular}
    \label{tab:smartphoneAppsList}
    \vspace{-4mm}
\end{table*}

\subsection{App Selection}
To examine how mobile AR applications request and present privacy permissions, we compiled a set of 23 popular Android apps (Table~\ref{tab:smartphoneAppsList}), each including at least one AR feature. These apps represent download counts ranging from 50,000 to 10~billion on the Google Play Store and span multiple categories, including social media, navigation, entertainment (e.g., gaming), and spatial measurement. We installed each app on an Android device, enabled its AR functionality, and documented all permission requests encountered.

Our documentation process followed the design space for privacy choices outlined by Feng et al.~\cite{FengEtAl}, capturing the specific permissions requested (e.g., camera, location, microphone), the timing of each request (e.g., first launch vs.\ just-in-time prompts), and the choice type (e.g., binary vs.\ multiple). For instance, some apps asked for camera access immediately upon launch, while others requested permission contextually when a particular AR feature was activated. Figure~\ref{fig:teaser} provides example screenshots illustrating how these prompts appear to users. By collecting these observations systematically, we gained a baseline for understanding how existing AR apps communicate their data requirements.

We developed an online survey to examine how users perceive AR-related permissions in greater detail. 
To keep the survey manageable, ensure response quality, and prevent participant fatigue, from the initial set of 23 apps, we selected at random 12 app-permission pairs that satisfied the following criteria: span the full range of permission  (i.e., camera, location, microphone, contacts, gallery, and phone calls), a mix of widely known apps (e.g. \emph{Google Maps}, \emph{Instagram}) and more niche apps (e.g. \emph{AR Ruler App}, \emph{Quiver}), and apps with obvious AR functionality (e.g. \emph{PokemonGo}) and apps with less obvious AR features (e.g. \emph{Google Maps}' Live View).

Each survey item first presented the one line app description and the first three official screenshots from the app’s listing on the Google Play Store (accessed April 2025). 
This structure emulates real-world scenarios in which users often install apps with limited information about data practices.
A subsequent question asked participants about their familiarity with the app, in order to capture prior exposure and contextualize their responses. 
Next, two 5-point Likert scale questions followed (ranging from ``very unlikely'' to ``very likely'') to determine how likely participants considered that the app would request a particular permission {and how likely it is that they would grant it if the permission was requested}. Immediately afterward, participants provided a brief open-ended explanation to justify their ratings. 
{The sample set of questions is provided in the Supplementary Materials.}

To ensure response variability, {9} items corresponded to actual permission requests identified in our app audit, whereas {3} were control items in which the specified permission was not actually requested by the app. These control items served as distractors to mitigate uniformly high likelihood ratings. 
The survey was built using \emph{SoSci Survey} and administered through \emph{Prolific}, gathering both quantitative (Likert-scale) and qualitative (open-ended) data. 

\subsection{Participants}
\begin{figure}[t]
    \centering
        \includegraphics[width=0.75\linewidth]{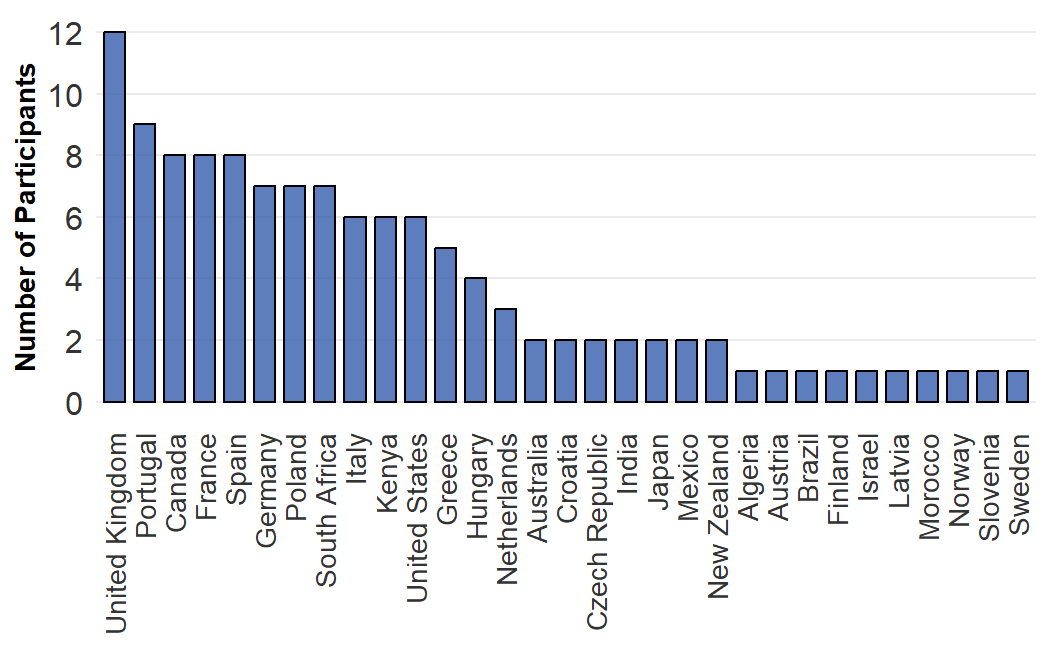}
    \caption{Distribution of participants by country of residence.}
    \label{fig:country}
\end{figure}

We deployed the survey on \emph{Prolific} in {April 2025}, recruiting {120} 
participants who met two criteria: fluency in English and regular smartphone usage. The final sample comprised {60 male and 60 female participants, aged 18 to 66 (M=34.51, SD=12.12).} 
{Participants represented a diverse international sample, with the highest numbers from the United Kingdom (12), and Portugal (9), followed by Canada, France and Spain (8 each). The distribution of participants by country of residence is shown in~\autoref{fig:country}.}
{Participants rated their technical proficiency on a scale from 1 (beginner) to 5 (expert) with a mean of 3.79 (SD=0.80). When asked about their prior experience with AR technology, a big majority (57) reported they are familiar with AR and have used it once or twice, 33 have used it more often, 27 are familiar but never used it, and 3 have not heard of it. Scores on the Internet Users’ Information Privacy Concerns (IUIPC) scale~\cite{malhotra2004internet} averaged 5.97 (SD 1.18) for Control, 6.36 (SD 1.11) for Awareness, and 5.54 (SD 1.46) for Collection.}
All participants received monetary compensation of {£10.20/hr} for their time. Before the study, we obtained ethical clearance from the Ethics Committee of LMU Munich.

\subsection{Data Analysis}

We analyzed the Likert-scale survey responses to identify overall patterns in user expectations and willingness to grant AR app permissions. Descriptive statistics (mean, standard deviation) were computed for each app--permission pair to asses perceived likelihood and consent behavior. We also examined control items to gauge response consistency and detect any potential systematic biases.

For the open-ended responses, we employed a thematic analysis to interpret participants' rationales. Two researchers initially coded a subset of the data (three app--permission pairs, accounting for approximately 13\% of all responses) and then refined the codebook through discussion. Three researchers subsequently used this consolidated codebook to code the remaining responses. Finally, the three researchers reviewed the complete coding results and clustered them into overarching groups and themes, yielding a qualitative perspective on user perceptions and misconceptions.
The final code book is provided in~\autoref{a_codebook}.

\subsection{Limitations}

Our findings are based on self-reported perceptions rather than observed in-situ behavior, which may not fully mirror real-world actions. While this approach enables the identification of user misconceptions—a critical goal for improving permission interfaces—it does not capture the immediacy of live app installations or just-in-time prompts. Additionally, our participant sample, though diverse in age and background, may not reflect the broader population of AR users. Future work can address these concerns by integrating field observations, usage logs, or longitudinal methods documenting how participants interact with AR permissions over time. 

\section{Results}
We report both quantitative and qualitative results on how users perceive privacy permissions in mobile AR. The quantitative outcomes from our Likert-scale items illustrate how likely participants expect certain permissions to be requested {and how likely it is that they would grant them should they be requested}. In contrast, the qualitative thematic analysis reveals how participants justify their responses, clarify (or conflate) permissions, and express misconceptions about AR data collection.

\begin{figure}[b]
    \centering
        \includegraphics[width=0.75\linewidth]{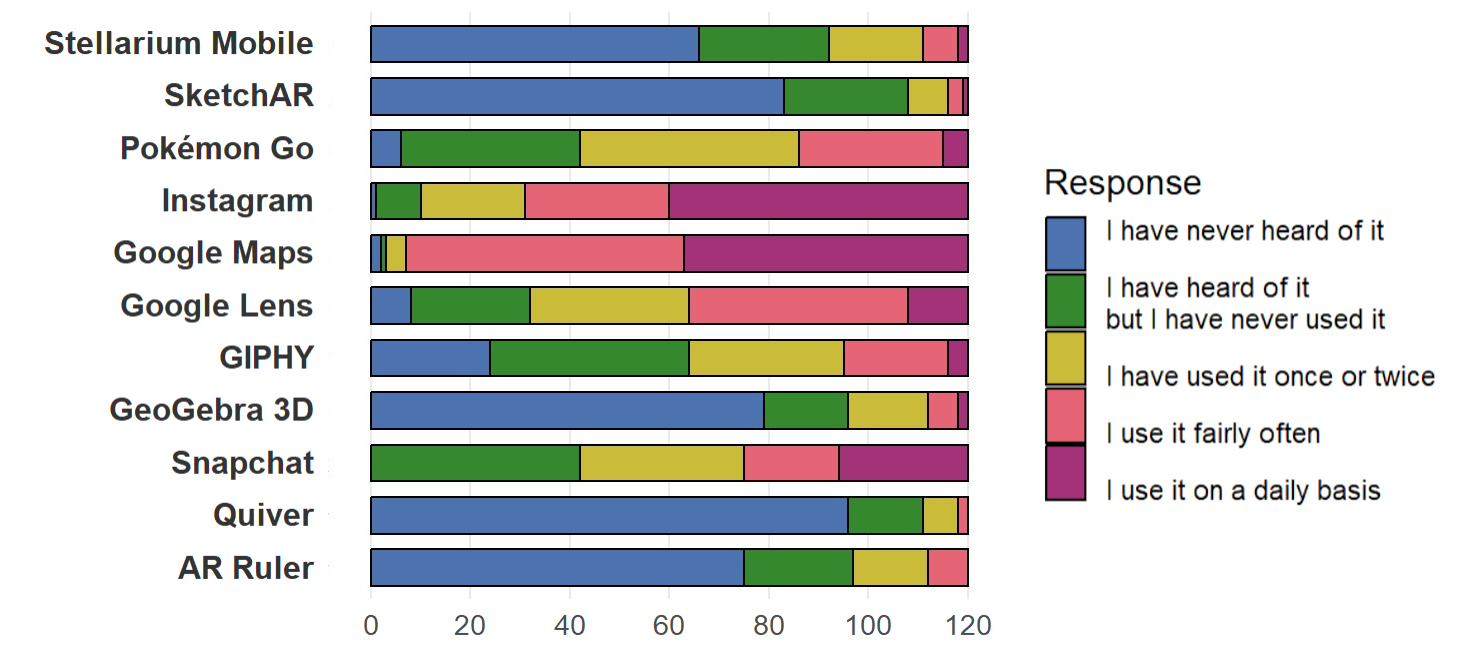}
    \caption{Self-reported participant familiarity with each app.}
    \label{fig:familiarity}
\end{figure}

\subsection{App Familiarity}

Participants’ familiarity with the selected apps varied widely, as shown in Figure~\ref{fig:familiarity}. Popular applications such as \emph{Instagram}, \emph{Snapchat}, and \emph{Google Maps} were well known, with many respondents reporting frequent or daily use. In contrast, niche AR applications such as \emph{AR Ruler}, \emph{SketchAR}, and \emph{Quiver} were unfamiliar to a large share of participants, the majority indicating that they had never heard of them. 

\subsection{Likelihood of Permission Requests and User Consent}

\begin{table*}[]
 \caption{  Summary of the results to the questions for likelihood that the app asks for the particular permission and the likelihood of the user to grant it answered on a 5-point Likert scale (1=Very unlikely to 5=Very likely).}
\footnotesize
   \centering
    \begin{tabular}{l  l l l}    
 \toprule
  \textbf{Application} &\textbf{Privacy Permission} & \parbox{4.8cm}{\textbf{Likelihood for App to Request Permission} \\ mean (SD)} & \parbox{4.8cm}{\textbf{Likelihood for User to Grant Permission} \\mean (SD)} \\ 
 \midrule
Instagram & location &  4.16 (1.12) & 3.24 (1.50)\\
Stellarium Mobile & location & 4.23 (0.99)  & 4.17 (1.05)\\
Snapchat  & calls &  3.37 (1.41) & 2.65 (1.60)\\
PokemonGo & contacts &  3.27 (1.28) & 2.48 (1.44)\\
GoogleLens &  gallery & 4.53 (0.83)  & 4.44 (0.90) \\
Quiver &  microphone & 2.18 (1.13) & 1.88 (1.15)  \\
GIPHY & camera &  4.11 (1.13) & 3.41 (1.33) \\
SketchAR & camera & 4.05 (1.00)  &  3.76 (1.22)\\
GeoGebra 3D & camera & 3.73 (1.41)  & 3.51 (1.37) \\
 \bottomrule
\end{tabular}
\label{table:MeanSD_new}
\vspace{-4mm}
\end{table*}

{Participants' responses revealed variations in both the perceived likelihood of apps requesting specific permissions and their own willingness to grant them (see Table~\ref{table:MeanSD_new}).

For some apps, such as \emph{Google Lens} (gallery) and \emph{Stellarium Mobile} (location), participants showed a strong alignment between expectation and willingness, with high mean scores for both (\emph{Google Lens}: 4.53 vs. 4.44; \emph{Stellarium Mobile}: 4.23 vs. 4.17). 
The justifications participants provided also showed that their understanding of why this permission was required for the functionality of the app was very high, which could explain the high acceptance rate.
However, in several other cases, users appeared more hesitant to grant permissions than they expected apps to request them. This was particularly evident for \emph{Instagram} (location) (4.16 vs. 3.24), \emph{Snapchat} (calls) (3.37 vs. 2.65), and \emph{Pokémon Go} (contacts) (3.27 vs. 2.48), suggesting a trust or necessity gap for more privacy-sensitive data types.
Answers also revealed that participants intended to use the apps without granting the permission by either only using other features of the app, or by using a workaround ("connect to other players via their unique identification number...no need to access contacts" P81 about adding friends in \emph{Pokémon Go} without granting access to contacts). Additionally, with giving access to contacts, participants pointed out that by doing so they would give away personal information of others, potentially raising a privacy concern ("I don't like the idea of sharing others' details with an app they haven't given permission to" P32).

Lower willingness scores for \emph{Quiver}’s microphone access (2.18 vs. 1.88) indicate both low expectation and low comfort with granting such access, likely due to unclear functional justification. Seven participants (P3, P19, P20, P40, P52, P80, P83) also mentioned that they were less comfortable with granting access to the microphone as the app was targeted for use with kids, indicating that they were more concerned for this vulnerable user group: "I would not really give it audio recording permissions, especially if kids are involved" (P83). In addition, not understanding for which feature the permission would be needed made some participants see the request as suspicious ("would be highly suspicious what is the use of the audio" P85). 

Meanwhile, camera-related permissions (e.g., \emph{GIPHY}, \emph{SketchAR}, \emph{GeoGebra 3D}) generally showed high or moderate expectations and slightly lower—but still relatively positive—willingness ratings, reflecting moderate user acceptance when AR functionality was perceived as plausible.
Participants further expressed that having a clear understanding of the feature helps them in making a decision, and makes it more likely that they would be willing to grant a permission ("I understand why it needs the location and so would be happy to allow permission" P14 on granting location access to the \emph{Stellarium Mobile} app; "I would want to know what they need it for" P100 on granting location to the \emph{Instagram} app). 
Looking at the likelihood ratings alone, participants generally showed strong expectations that mobile AR apps would request access to core sensor-based permissions. \emph{Google Lens} (gallery) received the highest likelihood rating (M = 4.53, SD = 0.83), reflecting a clear user understanding that the app relies on stored images for visual analysis. Similarly, \emph{Stellarium Mobile} (location) (M = 4.23) and \emph{Instagram} (location) (M = 4.16) were perceived as likely to request location data, likely due to their known use of geolocation for content tailoring and tagging. Camera-related permissions for \emph{GIPHY} (4.11) and \emph{SketchAR} (4.05) were also rated highly, suggesting users associate these apps with image capture or AR overlays. \emph{GeoGebra 3D} received a slightly lower likelihood score for camera access (3.73), potentially due to less awareness of its AR features. In contrast, apps requesting microphone (\emph{Quiver}: 2.18), calls (\emph{Snapchat}: 3.37), or contacts (\emph{Pokémon Go}: 3.27) scored notably lower, indicating some uncertainty or scepticism about the relevance of these permissions. These results suggest that while users expect certain “core” AR permissions like camera and gallery access, they are more cautious or uncertain about permissions perceived as peripheral or privacy-sensitive, especially when their connection to app functionality is less transparent.}

In some cases, the participants seemed unaware of a certain (AR) feature and thus indicated that they do not see a necessity for the app to request the permission in question. For example, with the \emph{Quiver~-- 3D Coloring App} that requests access to the device's microphone, participants who said it would not be needed ({P111: "I dont understand why a coloring app should ask for audio"}) are probably not aware that audio input could be used if you want to capture a video of the AR drawing. Similarly, with \emph{GeoGebra 3D}, the participants who indicated that camera access would not be needed for this application were likely unaware of the app's AR functionality. For example, participant {P72} stated, {"GeoGebra 3D Calculator is mainly a math and graphing tool, not something that typically needs camera access"}, which indicates that they did not know the camera could be used to position digital objects in the physical environment.

Independent of the functionality, participants also mentioned that either trusting or not trusting the app/app provider influenced whether they were comfortable granting a permission or not ("From a company that I know and trust so would 100\% share any details." P29). For apps that they were actively using, some participants also reported that they had already granted the permission in order to use the features relevant to them ("I already gave Snapchat this permission" P70).

Participants also reported that making use of Android's granular control would help them feel more comfortable by granting a permission only for one time or by restricting the usage to when the app is actively used ("I would have it setup to "ask permission each time" so that I could have some control" P44; "would probably restrict the location access to while the app is actively in operation" P2).



\begin{table*}[]
 \caption{ Summary of the results to the control questions for likelihood that the app asks for the particular permission and the likelihood of the user to grant it answered on a 5-point Likert scale (1=Very unlikely to 5=Very likely).}
\footnotesize
   \centering
    \begin{tabular}{l l l l}    
 \toprule
  \textbf{Application} & \textbf{Privacy Permission} & \parbox{5cm}{\textbf{Likelihood for App to Request Permission} \\ mean (SD)} & \parbox{5cm}{\textbf{Likelihood for User to Grant Permission} \\mean (SD)} \\ 
 \midrule
GoogleMaps & contacts &  2.88 (1.48) & 2.54 (1.54) \\
AR Ruler App & location &  2.59 (1.34) & 2.12 (1.29) \\
GeoGebra 3D & microphone & 1.93 (1.13)  & 1.89 (1.24) \\
 \bottomrule
\end{tabular}
\label{table:MeanSD_control}
\vspace{-4mm}
\end{table*}
The results from the control questions indicate that participants were generally less likely to expect these apps to request the specified permissions, and similarly less inclined to grant them (see Table~\ref{table:MeanSD_control}). \emph{GeoGebra 3D}’s microphone access, which was not actually required by the app, received the lowest expectation score (M = 1.93, SD = 1.13) and an almost identical willingness score (M = 1.89, SD = 1.24), suggesting participants correctly perceived the implausibility of this permission. This was also reflected in the explanations provided by the participants, for example P3 stated "It is unlikely that a calculator that scans or makes imagery would need audio."
For \emph{AR Ruler App} and location access, the expectation was slightly higher (M = 2.59, SD = 1.34), reflecting a common misconception identified in the study, that distance measurement tools require GPS. 
Particularly, 12 participants mentioned that they could imagine that the app "might need the GPS to do some measuring" (P73) and to "ensure measurements are accurate" (P112).
However, participants were somewhat less willing to grant this permission (M = 2.12, SD = 1.29), possibly due to privacy concerns or uncertainty about its relevance.
\emph{Google Maps} and contacts access had the highest scores among the control items, both in terms of perceived likelihood (M = 2.88, SD = 1.48) and willingness to grant (M = 2.54, SD = 1.54).
Some explanations participants thought of were that access to contacts could be used to "share addresses or track location with friends and family" (P3) or if "I want to drive to a friend it would look into the contacts and see if an address is linked to it" (P69).

\subsection{Experience Based Reasoning}
Participants also drew on personal experience with particular apps. In some cases, respondents recalled having previously granted a permission and thus confidently deemed it likely. Conversely, others did not remember ever encountering such requests and therefore considered the permission unlikely to be required. This firsthand recollection, or lack thereof, strongly influenced their estimates of permission necessity.

On a similar note, participants expressed that they expect an application to request a permission just because {"pretty much every app that I've had asks for permission to use the phone's camera" (P37)}. This was especially true for the permissions camera and location. {Additionally, some said they would simply expect it from the company which provides the app (e.g. Meta or Google), as stated for example by P65: "meta always wants to know where you are" or P83: "Google usually always asks for [this] permission".}

\subsection{Misconceptions}
The responses have also revealed misconceptions regarding some functions and the required permissions. 

Some responses to \emph{GIPHY} and \emph{Sketch AR} requesting camera access suggest that participants mixed up the functionality of accessing the camera or the photo gallery of the device. For \emph{Sketch AR}, some participants said the camera access would be needed to store sketches, which would need access to the gallery ({P72: "for saving or sharing the drawings."}).
Similarly, for \emph{GIPHY}, some participants thought the camera access request allowed the app to {use pictures from the gallery (P38: "I assume it can create a GIF/sticker based on your photo and therefore would need to access the gallery on your device"; P101: "Not sure if that is from gallery or live")}.

From the control questions, some reasons given to why \emph{AR Ruler App} would need access to the location revealed that participants might wrongly assume that the measurement functionality needs this permission to work ({P97: "I think it would be needed to measure a certain length", P112: "This would ensure measurements are accurate"}). 
The location permission, however, is not needed for the measuring to work, and the app does not actually request it.

\section{Discussion}
In this section, we interpret our survey results to identify key gaps between user expectations and actual permission practices in AR apps. Drawing on both Likert-scale responses and qualitative explanations, we highlight five major themes.

\subsection{{Contextual Understanding Drives Trust and Consent}}
{Participants showed a strong willingness to grant permissions when they perceived a clear and direct connection between the permission request and the application's AR functionality. For instance, applications like \emph{Google Lens} (gallery access) and \emph{Stellarium Mobile} (location access) received high ratings not only in terms of expected permission requests but also in users’ stated willingness to approve them. Participants often explained their decisions by referencing specific app features they associated with the requested data, such as image analysis or geolocation-based stargazing.}

{
This alignment highlights the role of functional transparency in supporting user trust. When the purpose of a permission is intuitive and clearly related to visible features, users are more likely to consider it legitimate and necessary. Conversely, abstract or generic permission labels without sufficient explanation can lead to confusion or hesitation, even when the data might be technically needed for AR functionality.}

{
To bridge this gap, design strategies such as contextual, just-in-time permission prompts—delivered at the moment when a feature is accessed—can make data use more understandable. Pairing these prompts with brief, task-specific explanations (e.g., “Gallery access enables you to import a reference image for AR analysis”) helps users align system requests with their mental models of the app’s operation.}

\summaryBox{}{Implement context-aware permission requests that clearly explain how specific data supports core AR functionality at the time it is needed, enhancing user comprehension and consent accuracy.}

\subsection{{Resistance to Peripheral and Sensitive Permissions}}

{
Participants consistently showed lower willingness to grant permissions perceived as peripheral or unrelated to core AR functionality, such as access to contacts, phone calls, or the microphone.
For example, while \emph{Snapchat}’s request for call access and \emph{Pokémon Go}’s request for contact access were moderately expected, many participants were significantly less inclined to grant them. Justifications commonly cited the perceived irrelevance of these permissions to the primary function of the app, as well as concerns about third-party privacy.
Also, participants mentioned that instead of granting these permissions they would rather use an alternative mechanisms already supported by the platform or their social networks.
For instance, instead of importing contacts, participants preferred to connect with friends in \emph{Pokémon Go} using the Player ID system provided in-app. Similarly, rather than enabling call permissions in \emph{Snapchat}, participants indicated they would initiate voice communication through other apps.}

{
This hesitancy was even more pronounced in the case of \emph{Quiver~-- 3D Coloring App}’s request for microphone access, which received the lowest overall willingness score. Participants expressed confusion as to why a children's coloring app would require audio input, especially in the absence of visible voice-based features.
}

\summaryBox{}{Avoid requesting sensitive permissions unless they are clearly necessary and well-explained. Where possible, offer privacy-preserving alternatives, and avoid placing core functionality behind access to data whose relevance is unclear, particularly in applications intended for or used by vulnerable user groups.
}

\subsection{Misconceptions About AR Distance Measurement and Location Access}
Our data revealed repeated confusion between depth sensing and geolocation. {12} participants believed that measuring objects or rooms in AR depended on GPS rather/in addition to the device camera, implying an oversimplified view of how AR processes space. This misconception may reflect a broader lack of familiarity with 3D sensor fusion, with some participants equating ``distance measurement'' to location tracking on a map.

Such misunderstandings can affect user trust and willingness to grant permissions. Suppose users mistakenly believe an app tracks their geographic whereabouts. In that case, they might deny access—even if the app actually uses camera-based techniques that carry different (and potentially lower) privacy risks.

\summaryBox{}{Use targeted in-app explanations clarifying the technological basis of specific AR features (e.g., ``This measuring tool uses your camera, not your location, to gauge distances''). Providing this detail when permission requests appear can dispel misconceptions that might otherwise lead to unnecessary refusals.}

\subsection{Confusion Between Similar Permissions}
Another common theme in participant responses was the conflation of permissions that serve related but distinct purposes, such as camera vs. gallery access. Multiple participants assumed that once an app could take new pictures or record video, it also had full access to stored images. This confusion not only raises privacy concerns, implying broader data reach than intended, but can also spark user suspicions about the motivations of the app.

Comments about \emph{GIPHY} illustrate this issue: although capturing new GIFs logically involves camera permission, some participants mistakenly believed that this also enabled the app to access their existing media library. 

\summaryBox{}{Distinguish between capturing new media and accessing existing media. Instead of generic labels, prompts should clearly convey their specific purpose and scope. For example: “Allow this app to use your camera to take pictures or videos?” vs. “Allow this app to access your photo library to select stored images?” 
These distinctions can be further reinforced through brief, in-context explanations (e.g., “Used to overlay AR effects on your saved images.”).
}

\subsection{Contextual Explanations to Improve Permission Understanding}
Several user misunderstandings traced back to insufficient context around why a particular permission is needed. \emph{SketchAR}, for instance, prompted participants to grant camera access for its AR-based drawing assistant, but some users, unfamiliar with how the feature worked, questioned the need for camera data. Similarly, apps like \emph{Quiver} confused participants who were unable to see how microphone access related to interactive coloring activities.

Providing explicit, just-in-time explanations can help bridge this gap. Applications can illustrate how each permission aligns with user-facing features. Such explanations reduce the guesswork, leading participants to distrust or dismiss permission requests as overreaching.

\summaryBox{}{Integrate contextual permission prompts triggered by specific feature engagement. For instance, when a user activates an AR drawing guide, a concise message (``We need camera access to project drawing templates onto your paper in real time.'') can clarify intent, preempt doubts, and foster greater acceptance.}

\subsection{Granular Permission Controls to Address Misconceptions}
Participants sometimes overestimated an app’s dependence on certain permissions. For instance, some believed \emph{Pokémon Go} needed contact access to enable friend-based gameplay—even though this feature is optional and primarily used to scan the local address book. Players could still add their friends by using their in-game ID without having to grant \emph{Pokémon Go} access to their contacts. Bundling such an optional feature into the core permission set can produce confusion and deter user adoption of social components.

Granular permission controls that allow users to select which features to enable can mitigate confusion and encourage a sense of autonomy. Users who only want the baseline AR functionality do not have to grant social or auxiliary permissions, while those seeking the full experience can opt-in voluntarily.

\summaryBox{}{Implement modular permission requests separating core AR functionality from auxiliary or social features. Clearly label permissions as “optional” where possible, and explain the added benefits for users who choose to grant them. This approach fosters informed consent and lets users decide which facets of the experience align with their comfort level.}

\section{Future Work}
While our study provided important insights into user perceptions of privacy permissions in mobile AR, several opportunities remain for extending this research. First, our survey-based approach focused on self-reported attitudes rather than observed behavior. Future work should incorporate longitudinal or in-situ methods to examine how users interact with permission prompts in real-world settings, especially during first-time use or after updates that introduce new AR features.

Second, although we captured a diverse international sample, user understanding of privacy may differ significantly across cultural, regulatory, and linguistic contexts. Cross-cultural comparative studies could reveal how local norms, legal frameworks (e.g., GDPR, CCPA), or regional mobile ecosystems influence expectations and consent behavior in AR environments.

Third, our findings identified recurring misconceptions about AR features embedded in apps that are not explicitly marketed as AR. Future research should explore how interface cues, iconography, or onboarding sequences can effectively signal AR functionality and its privacy implications to users, particularly those with limited technical background.

Lastly, as wearable AR devices (e.g., glasses) gain traction, privacy concerns become even more pronounced, given the continuous presence of body-proximal sensors \cite{nair2023exploring,nair2023uniqueID,paneva2024ieeepvc}. Capturing bystander data, whether intentionally or not, highlights the complexity of managing multi-stakeholder consent and safeguarding user boundaries. Beyond the primary user, bystanders and non-users often remain unaware of how they might be recorded or analyzed in AR.

Finally, evolving multi-device or mixed-reality ecosystems bring additional challenges for cross-platform permission management. As individuals move seamlessly between physical and digital realms, next-generation permission frameworks will need to adapt, ensuring consistent privacy protections even when multiple devices or applications handle overlapping data streams. Developing robust, context-aware solutions that address these multifaceted scenarios stands as a critical goal for future AR privacy research.

\section{Conclusion}
Our study analyzed user perceptions and misconceptions surrounding privacy permissions in mobile AR applications. 
By examining real-world permission requests by some of the most downloaded Android AR apps and surveying 120 participants, we provide empirical evidence to inform the development of AR-specific privacy solutions. 
Addressing these issues through clearer permission prompts, modular controls, and contextual explanations can enhance user trust and transparency, leading to more effective privacy frameworks for mobile AR technologies.
These findings highlight the need for more user-friendly privacy mechanisms that cater to varying levels of technical literacy. Future mobile AR design should prioritize intuitive and contextual permission interfaces to mitigate user confusion. 

\begin{acks}
This work has received funding from the German Research Foundation (DFG) under grant agreement no. 521584224.
The authors thank all participants in the online study.
\end{acks}

\bibliographystyle{ACM-Reference-Format}
\bibliography{sample-base}

\appendix
\section{Code Book}
\label{a_codebook}

Final coding tree for the thematic analysis:
\begin{itemize}
    \item Functionality
        \begin{itemize}
        \item Feature Necessary 
        \item Feature Not Necessary 
        \item Optional Feature / Enhancing User Experience 
        \item Data Required for AR Features
        \item Plausible / Expected Permission Request
        \item Functionality Unclear
        \item Perceived Benefit of Feature 
        \item No Perceived Benefit of Feature
        \item Misconception (Regarding the Functionality Behind the Permission Request)
        \end{itemize}
    \item Decision
        \begin{itemize}
            \item Factors Influencing Decision
                \begin{itemize}
                    \item Trust
                    \item Prior Experience 
                    \item Privacy Concerns
                \end{itemize}
            \item Decision Outcome / Willingness to Grant Permission
                \begin{itemize}
                    \item Willing
                    \item Not Willing
                    \item Does Not Have a Choice
                    \item Willing to Grant Limited / Granular Access
                    \item Willing to Grant only if a Condition is Met
                \end{itemize}
        \end{itemize}
    \item Likeliness for the App to Request Permission
                \begin{itemize}
                    \item Likely
                    \item Unlikely
                    \item Must Ask
                \end{itemize}
    \item Unsure / No Answer
    \item Other
\end{itemize}

\end{document}